\documentclass[a4paper]{jpconf}
\usepackage{epsfig,amssymb,xspace,graphicx}
     
     \newcommand{\pathnow}{}
\font\twelvegoth=eufm10 at 14.4pt
\newfam\gothfam
\textfont\gothfam=\twelvegoth

\font\twelveron=eusm10 at 14.4pt
\newfam\ronfam
\textfont\ronfam=\twelveron

\def\lessim{\lower.5ex\hbox{$\; \buildrel < \over \sim \;$}}
\def\gtrsim{\lower.5ex\hbox{$\; \buildrel > \over \sim \;$}}

\newcommand{\nc}{\newcommand}
\nc{\ds}{\displaystyle}        \nc{\ts}{\textstyle}
\nc{\rf}[1]{Fig.\,\ref{#1}}    \nc{\rt}[1]{table\,\ref{#1}}
\nc{\req}[1]{Eq.\,(\ref{#1})}  \nc{\eps}{\varepsilon}
\nc{\beq}{\begin{equation}}     \nc{\beql}[1]{\begin{equation}\label{#1}}
\nc{\eeq}{\end{equation}}        
\nc{\beqa}{\begin{eqnarray}}   \nc{\eeqa}{\end{eqnarray}}       
\nc{\bfi}{\begin{figure}}       \nc{\efi}{\end{figure}}

\nc{\clbox}[1]{

\begin{center}\framebox{#1}\end{center}
}
\nc{\flbox}[1]{\fboxrule=5pt\framebox{#1}\fboxrule=2pt}
\begin{document}
\title{Strangeness and the Discovery of\\ Quark-Gluon Plasma}

\author{Johann Rafelski$^1$, Jean Letessier$^2$}

\address{$^1$Department of Physics, University of Arizona, TUCSON, AZ 85718, USA%
}%
\ead{Rafelski@Physics.Arizona.EDU}

\address{$^2$Laboratoire de Physique Th\'eorique et Hautes Energies\\
Universit\'e Paris 7, 2 place Jussieu, 75251 Cedex 05, France}%
\ead{JLetes@LPTHE.Jussieu.FR\\}

\begin{abstract}
 Strangeness flavor yield $s$ and the entropy yield $S$
are  the observables of  
the  deconfined quark-gluon state of matter  which can be 
studied in the entire available experimental energy range at AGS, SPS, 
RHIC, and, in near future,  at the LHC  energy range. We present 
here a comprehensive
analysis  of strange, soft hadron production 
as function of energy and reaction volume.  We discuss the 
physical properties of the final state and argue how evidence about the
primordial QGP  emerges. 
\end{abstract}.

\vskip -12.cm\noindent 
{\it 5th International Conference on Physics and Astrophysics of Quark Gluon Plasma,
}\\ February 8 - 12, 2005,  Salt Lake City, Kolkata, India, \\
{\it Journal of Physics: Conference Series}
\vskip 11.cm
\section{Introduction}\label{intro}
The deconfined interacting  quark--gluon plasma phase
(QGP) is the {\it equilibrium} state of matter at high temperature
and/or density.  It is   believed that  this state has  been 
present in the early Universe, 10-20$\mu$s into 
its evolution. The  question is if, in the short
time, $10^{-22}$--$10^{-23}$~s,  available in a laboratory heavy ion
collision  experiment, the color frozen nuclear phase 
can melt and turn into the QGP state of matter. There is
no valid first principles answer to this question 
available today, nor as it seems,
will a first principles simulation of the dynamic heavy ion
environment become available in the foreseeable   future. 
To   address this  issue we study  QGP experimentally,
which requires  development of laboratory experiments and 
suitable  observables. 

To form QGP in the laboratory   we perform  relativistic 
heavy ion collisions in which a domain of (space, time) 
much larger than normal hadron  size is formed, 
in which color-charged  quarks and gluons are
propagating  constrained by external 
`frozen  vacuum', which abhors color~\cite{RBRC}.
 We expect a pronounced  boundary in temperature and baryon density
between confined and deconfined  phases of matter, irrespective 
of the question if  there is, or not, a true phase transition. 
We search for a boundary between phases considering   the 
size of the interacting region and the magnitude of the reaction energy. 
 
Detailed study of the properties of the deconfined state
shows that QGP is rich in entropy and strangeness. 
The enhancement of entropy $S$ arises
because the color bonds are broken and gluons can be created. 
Enhancement of  strangeness $s$ arises  because the 
mass threshold for strangeness excitation is considerably lower 
in QGP than in hadron matter. Moreover there are 
new mechanisms of strangeness formation in QGP involving 
reactions between (thermal) gluons. Thus $S$ and $s$ 
 are the two elementary observables
which are  explored with soft hadronic probes, for
further theoretical details and historical developments
see our book~\cite{CUP}.  The 
numerical work presented here was carried out with the 
public package of programs SHARE~\cite{share}. This 
report is a  self contained summary of   our recent results,
 see~\cite{AGS,bdepend,edepend,acta03}.
 
Entropy enhancement, observed in terms of enhanced hadron multiplicity per 
net charge, has been among  the first indications of 
new physics reach of CERN-SPS experimental heavy ion program~\cite{entro}. 
The enhancement of strange hadron 
production  both as function of  the   number of  participating baryons, 
and reaction energy has been explored in several experiments at at BNL-AGS, 
CERN-SPS and and BNL-RHIC. 
We refrain from extensive historical survey of these results and present 
perhaps the latest, STAR-RHIC result in \rf{RHICEnhance}~\cite{Caines:2004ej}. 
In this presentation one sees the yield per participant  $N_{\rm part}$ divided 
by a reference yield obtained in $pp$ reactions. We observe  that 
the enhancement rises both with the strangeness content in  the hadron and 
with the size of the reaction region, indicating that the cause of this 
enhancement is a increased yield of strange quarks in the source, a qualitative
 expectation  we will address in our quantitative analysis below. The gradual 
increase of the enhancement over the range of $N_{\rm part}$ is an important
indicator of the physics mechanisms at work. This behavior agrees with our
studies of kinetic strangeness production and strangeness yield increasing 
with the size of the reaction region. This  enhancement  of strange 
antibaryons which  demonstrates that a novel  strangeness  production  
mechanism is present has been extensively studied at SPS energy range, 
where it was originally discovered~\cite{Bruno:2004pv,Gaz}.  

\begin{figure}[htb]
\vskip 0.2cm
\centerline{
\psfig{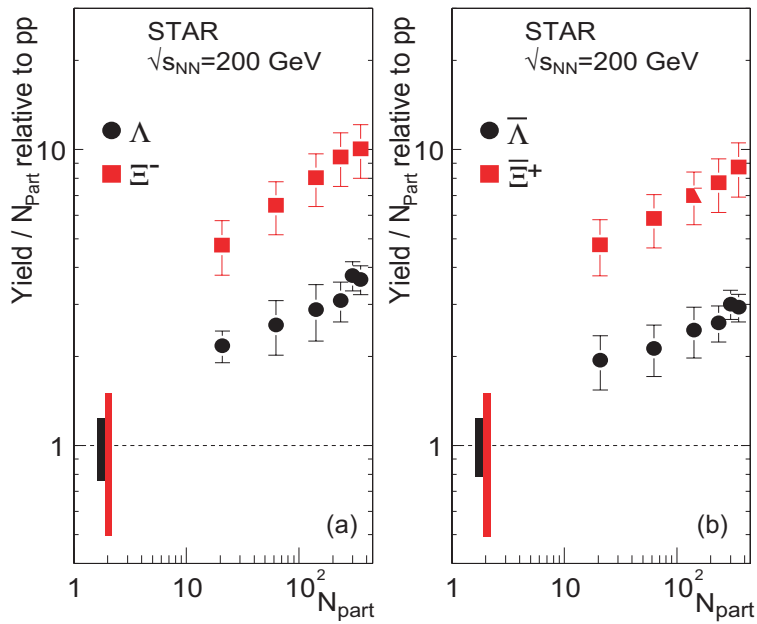}
}\vspace*{-0.3cm}
\caption{
\label{RHICEnhance} 
Yields  per participant $N_{\rm part}$ relative to $pp$  of 
 $\Lambda$ and $\Xi^{-}$ on left  and  $\bar{\Lambda}$ and $\bar{\Xi}^{+}$
on right   in Au+Au collisions at $\sqrt{s_{\rm NN}}=200$\,GeV 
Error bars  are statistical.
Ranges for $pp$ reference data  at $N_{\rm part}=2$  indicate the systematic
uncertainty.
}
\end{figure}

Further evidence for parton dynamics  prior to final state
hadronization is obtained from the study of strange hadron  
 transverse energy spectrum.  
The  identity of hyperon and antihyperon spectra 
in particular  $\Lambda, \overline\Lambda$ 
and    $\Xi, \overline\Xi$ implies 
that both strange matter and  antimatter must have been
 produced from a common source  by the same  
 fundamental mechanism. Furthermore they were not   subject to 
interactions  in their passage through the
baryon rich hadron gas present at the  SPS energy range. 

The  $\overline\Lambda, \overline\Xi$ annihilation in baryon-rich hadron gas 
 is strongly momentum dependent. This should  deform  
the shape of the  antihyperon spectra as compared to the 
spectra of hyperons. Thus symmetry of the hyperon-antihyperon spectra
also implies that there was no appreciable annihilation  of 
the $\overline\Lambda,\overline\Xi$  after their formation.  
The  working hypothesis is therefore
 that hadronization of the QGP deconfined phase formed
in high energy nuclear collision is direct, fast (sudden) and 
occurs without significant sequel interactions.  
That can be further tested in a  study of yields 
and spectra of unstable resonances~\cite{Rafelski:2001hp}. 

At RHIC the parton level dynamics is convincingly demonstrated
by quark content scaling of azimuthal asymmetry of the collective
flow $v_2$. Further evidence is derived from the consideration
of quark recombination formation of hadrons from QGP. We refer to
the recent comprehensive survey of the RHIC result~\cite{RBRC}, and   
presentations at the meeting addressing these very interesting and
recent developments. We will assume in this report that the 
case of quark-parton dynamics prior to hadronization is 
convincing, and will use our analysis of soft hadron 
production to find the  thresholds of the onset of deconfinement 
and to determine the properties of the deconfined fireball at
the time of its breakup into hadrons. 

In  next section \ref{StatHad} we introduce the 
statistical hadronization method  of analysis
of hadron production. We discuss   data analysis as 
function of  impact parameter and energy dependence of
soft hadron production
in section \ref{depen}. We will present both the systematics
of statistical model parameters, and the associated
physical properties. In section \ref{compare}, we address
the physical QGP signatures indicating presence of a phase 
boundary, giving particular attention to the explanation of 
the `horn' in the ${\rm K}^+/\pi^+$ ratio, and the   
strangeness to entropy relative yield. We discuss,
in section \ref{final}, the role these results play in 
understanding of the phase boundary separating QGP 
from normal confined matter. We also discuss briefly possible
new soft hadron physics at LHC.

\section{Statistical hadronization model }\label{StatHad}
To describe the yields of particle produced we employ
the statistical hadronization model (SHM). SHM  is by definition a model of 
particle production in which the birth process of each particle 
fully saturates (maximizes) the quantum mechanical probability amplitude, and
thus, the relative  yields are determined solely 
by the appropriate integrals of the
accessible phase space.
 For a system subject to global dynamical 
evolution such as collective flow,  this is understood to apply 
within each local  co-moving frame.
 
When particles are produced in hadronization, 
we speak of chemical freeze-out.  Hadron formation from QGP phase has 
to absorb the high
entropy  content of QGP which originates in broken color bonds.
The lightest hadron is pion and most entropy per energy 
is consumed in hadronization  by producing these
particles abundantly. It is thus important to 
free the yield of these particles from the chemical 
equilibrium constraint. 

The normalization of the   particle yields is,  aside of
the freeze-out temperature $T$,
directly  controlled by the particle fugacity 
$\Upsilon_i\equiv e^{ \sigma_i  /T}$, where $ \sigma_i$
 is particle `$i$' chemical potential. Since for each related  
particle and antiparticle  pair, we need two chemical potentials, it 
has become convenient to choose parameters such that we can control 
the difference, and sum of these separately.  For example, for nucleons $N$,
and, respectively,  antinucleons $\overline{N}$ the
two chemical factors are: 
\begin{equation}
\sigma_{N}\equiv \mu_b +T\ln\gamma_N ,\qquad
\sigma_{\overline{N}}\equiv -\mu_b +T\ln\gamma_N,
\end{equation}
\begin{equation}
\Upsilon_N=\gamma_N e^{ \mu_b /T}, \qquad\qquad
\Upsilon_{\overline{N}}=\gamma_N  e^{- \mu_b /T}.
\end{equation}

The   (baryo)chemical potential 
 $\mu_b$,  controls the baryon number, arising from the particle difference. 
$\gamma_N$, the phase space occupancy,   regulates the number of nucleon--antinucleon pairs present. 
There are many different hadrons, and in principle, we could
assign to each a chemical potential and then look for 
chemical reactions which relate these chemical potentials. 
However, a more direct way to accomplish
the same objective consists in  characterizing 
each particle by the valance quark content.
The relation between quark based fugacity and  chemical potentials 
($\lambda_{q,s}=e^{\mu_{q.,s}/T}$) and the
two principal  hadron based chemical potentials of baryon number and 
hadron strangeness $\mu_i,\  i=b,{\rm S}$ is:
\beql{muhadq}
\mu_b=3\mu_q\qquad
\mu_s=\frac 1 3 \mu_b-\mu_{\rm S},
\qquad \lambda_s={\lambda_q\over\lambda_{\rm S}}.
\eeq

An important (historical)  anomaly is the negative 
S-strangeness in $s$-carrying-baryons.
We will in general follow quark flavor and use quark chemical factors
to minimize the confusion arising. 
In the local rest frame, the  particle yields  are 
proportional to the momentum integrals of the quantum  distribution. 
As example, for the yield of pions $\pi$, nucleons $N$ and antinucleons 
$\overline N$ we have:
\begin{eqnarray}\label{Npi}
\pi &=& {V} g_\pi\!\!\int\!\!\frac{d^3p}{(2\pi)^3}
  \frac{1}{\gamma_q^{ -2}e^{E_\pi/T}-1}\,,
     \qquad   E_i=\sqrt{m_i^2+p^2},\quad \gamma_q^2<e^{m_\pi/T}\\[0.3cm]
N&=&{V}g_N\!\!\int\!\!\frac{d^3p}{(2\pi)^3}
   \frac{1}{\gamma_q^{ -3}\lambda_q^{ -3 }e^{E_N/T}+1},\quad
\overline {N}=  {V}g_N\!\!\int\!\!\frac{d^3p}{(2\pi)^3}
    \frac{1}{\gamma_q^{ -3 }\lambda_q^{ +3 }e^{E_{\bar N}/T}+1}.
\end{eqnarray}
 
There are  two types
 of chemical factors $\gamma_i$ and $\mu_i$, and thus
two types of chemical equilibriums. These are shown 
 in table~\ref{parameters}. The absolute
equilibrium is reached when the phase space occupancy approaches unity,
$\gamma_i\to 1$. The distribution of flavor (strangeness) among many
hadrons is governed by the relative chemical equilibrium. 

\begin{table}[hbt]
\caption{\label{parameters}Four quarks $s,\ \overline{s},\ q,$ and $\ \overline{q} $
 require four chemical parameters; right: name of the associated chemical equilibrium} 
\vskip 0.3cm
\begin{center}
\begin{tabular}{ll|l}
\hline
\hline 
 $\gamma_{i}$&   controls overall abundance & Absolute   chemical\\
&of quark  ($i=q,s$)  pairs & equilibrium\\
\hline
 $\lambda_{i}$&   controls difference between & Relative   chemical\\
&quarks and antiquarks  ($i=q,s$) & equilibrium
\end{tabular}
\end{center}
\end{table}

In order to arrive at the full particle yield, 
one has to be sure to include all the hadronic resonances
which decay feeding into the  yield considered, {\it e.g.}, the decay 
 $K^*\to K+\pi$ feeds into $K$ and $\pi$ yields. The contribution
is sensitive to temperature at which these particles are formed. 
Inclusion of the numerous resonances   constitutes a book 
keeping challenge in study of particle 
multiplicities, since decays are contributing at the 50\% level to 
practically all particle yields. A public statistical
hadronization program, SHARE (Statistical HAdronization with REsonances)
has simplified this task considerably \cite{share}.

The  resonance decay contribution is
dominant for the case for the  pion yield. This happens  
even though each resonance contributes relatively 
little in the final count. However,  
the large number of resonances which  contribute compensates 
and the sum of small contributions   competes with the 
direct pion yield. For the more heavy hadrons, generally
 there is a dominant  contribution from just a few, or even
from a single  resonance. The exception are the $\Omega,\overline\Omega$ 
which have no known low mass resonances,  and also $\phi$ -- the known
resonances are very heavy and very few.

A straightforward   test  of  sudden hadronization and the  SHM   is that 
within a particle `family',  particle yields with same valance quark
content are in relation to each other well described by integrals 
of relativistic phase space. The relative yield of, {\it e.g.},
$K^*(\bar s q)$ and $K(\bar s q)$ or $\Delta$ and $N$  are  controlled 
by the particle masses $m_i$,  statistical weights (degeneracy) $g_i$ and the 
hadronization temperature $T$. In the Boltzmann limit
one has (star denotes the resonance): 
\begin{equation}\label{RRes}
{N^*\over N}= {g^*m^{*\,2}K_2(m^*/T)\over g\,m^{2}K_2(m/T)}.
\end{equation}
Validity of this relation implies insensitivity of the quantum matrix element 
governing the coalescence-fragmentation production of particles to
intrinsic structure (parity, spin, isospin), and  particle mass. 
The measurement of the relative yield of hadron resonances 
is a sensitive test  of the statistical hadronization hypothesis
and lays the foundation  to the application of the method in data analysis. 

The method available to measure resonance yields 
depends in its accuracy significantly on the   
the precise nature of   the hadronization process:
the resonance yield is derived  by reconstruction of the invariant mass of the 
resonance from decay products energies $E_i$ and
momenta $p_i$. Should  the decay products of resonances  
after formation  rescatter on  other particles, then often 
their energies and momenta will change enough for the   invariant mass
to fail the    acceptance conditions set in the experimental
analysis.  Generally,   
the rescattering effect depletes more strongly the yields of 
shorter lived resonances , as a greater fraction of these will
decay shortly after formation, when elastic   scattering of decay 
products on other produced particles  is possible. 

We further hear often  the argument   
  that the general scattering process of hadrons in matter
can form additional resonance
states. In our opinion, the   loss of observability (caused by {\it any}
scattering of {\it any} of the decay products is considerably greater
than a possible production gain. The loss of resonance yield 
 provides additional   valuable  information about the freeze-out 
conditions (temperature and time duration)\cite{Rafelski:2001hp}.

\section{Phase thresholds: volume and energy dependence}  \label{depen}
\subsection{Statistical model parameters}\label{paramet}
In order to explore the properties of the fireball at hadronization
as function of the volume   at
 the top RHIC energy $\sqrt{s_{NN}}=200$\,GeV Au--Au, we study
 the 11 centrality bins in which the 
$\pi^\pm, {\rm K}^\pm, p$ and $\bar p$ rapidity yields $dN/dy$ for 
$y_{\rm CM}=0$ have been recently 
presented, see table I and table VIII in Ref. \cite{phenixyield}. 
These 6 particle yields and their ratios change rapidly. 
On the other hand, the additional two experimental
results, the    STAR  
${\rm K}^*(892)/{\rm K}^-$ \cite{haibin2200},
and  $\phi/{\rm K}^-$  \cite{phiyld}  show 
little centrality dependence.

In addition, three  supplemental constraints help to determine the best fit:\\
A) strangeness conservation, {\it i.e.\/}, the (grand canonical)
 count of $s$ quarks in all hadrons equals such  $\bar s$ count for 
each rapidity unit;\\
B) the electrical charge to net baryon ratio 
in the final state is the same as in the initial state;\\
C) the ratio   $\pi^+/\pi^-=1.\pm0.02$, which helps constrain the 
isospin asymmetry.\\
This last ratio appears redundant, as we already independently  use
the yields of $\pi^+$ and $\pi^-$. These yields have a large systematic
error and do not constrain their ratio well,
 and thus the supplemental constraint
is introduced, since SHARE allows for the isospin asymmetry effect.
The  7 SHM parameters (volume per unit of rapidity 
 $dV/dy$, temperature $T$, four chemical 
parameters $\lambda_q, \lambda_s, \gamma_q, \gamma_s$ and the 
isospin factor $\lambda_{I3}$ are in this case
studied in  a systematic fashion as function of impact parameter using 
11 yields and/or ratios and/or constraints, containing one (pion ratio) 
redundancy. 

Although the number of degrees of freedom in such 
analysis is small, the  $\chi^2$ minimization  yielding good
significance is easily accomplished, showing good consistency of the
data sample. The resulting statistical parameters are shown
in \rf{gammu}, as function of participant number. 
We show on left results for the full  non-equilibrium model allowing 
  $\gamma_q\ne 1, \gamma_s\ne 1$ (full circles, blue) and semi-equilibrium setting
$ \gamma_s= 1$ (open circles, red). From top to bottom the  (chemical)
freeze-out temperature $T$, the occupancy factors 
$\gamma_q$, $\gamma_s/\gamma_q$ and together in the
bottom panel  the baryon   $ \mu_B$ and hyperon $\mu_S$ 
chemical potentials. On left, in \rf{gammu}, we present the results for the 
centrality dependence, as function of participant number,
on right as function of energy.

\begin{figure}[!bt]
\hspace*{-.1cm}
 \psfig{width=8.5cm,figure=\pathnow 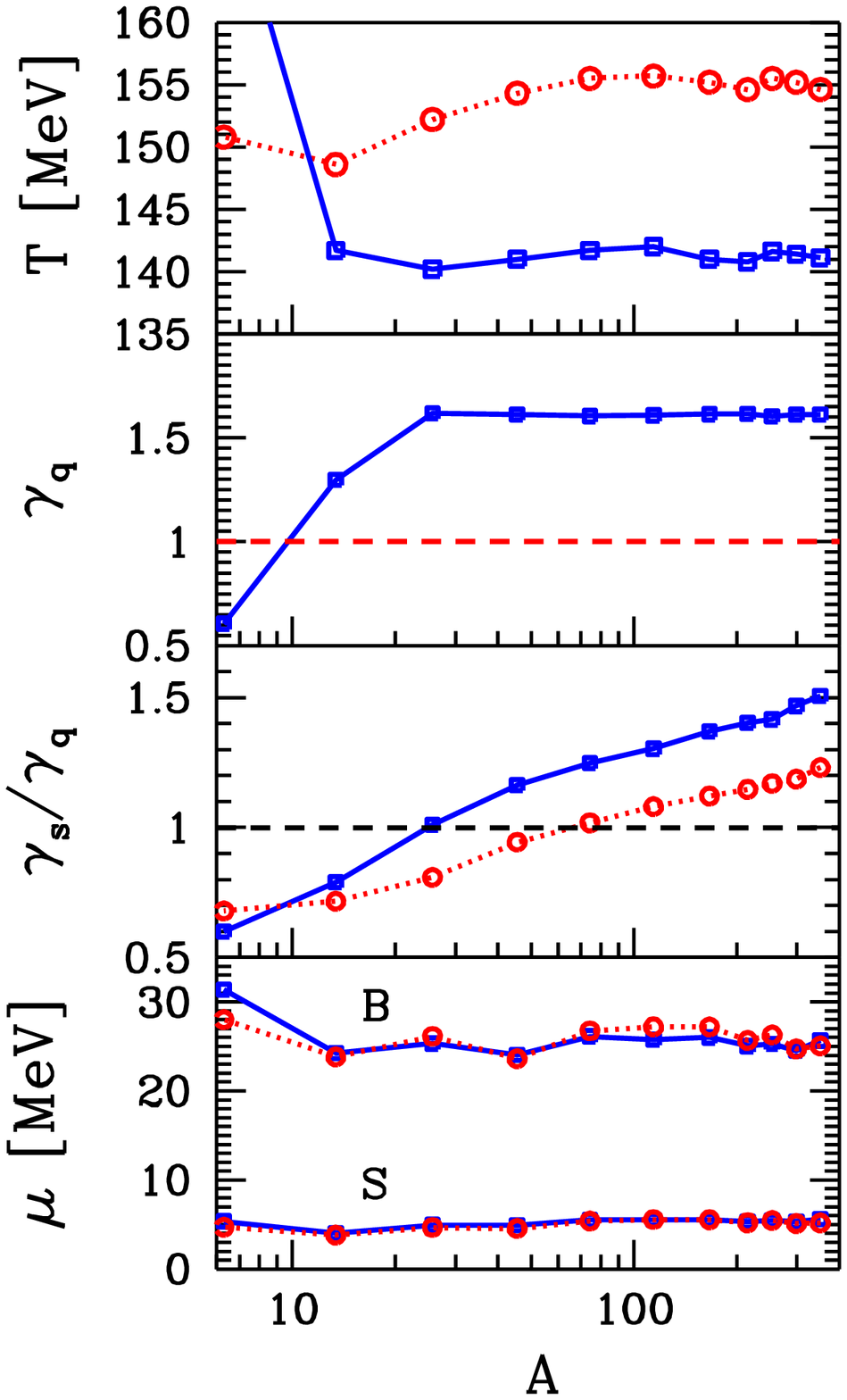} 
 \hspace*{-.50cm}\psfig{width=8.5cm,figure=\pathnow    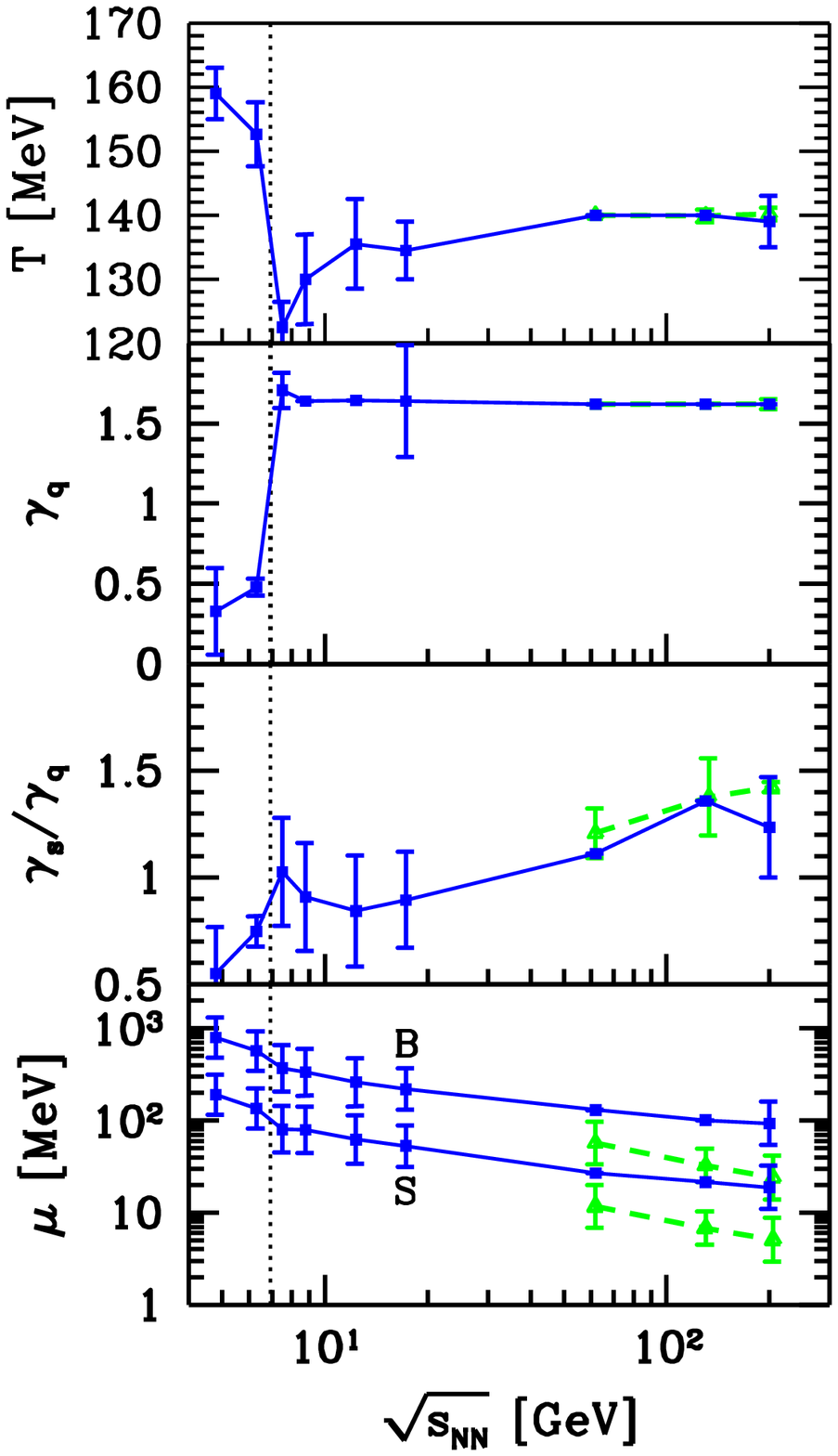}
 \caption{ \label{gammu} 
 From top to bottom: temperature $T$, light quark phase 
space occupancy $\gamma_q$,
the ratio  of strange to light quark phase 
space occupancies $\gamma_s/\gamma_q$  and the chemical potentials
($B$ for baryochemical $\mu_B$ and $S$ for strangeness  $\mu_S$) 
The lines
guide the eye.
Left:  
as function of centrality {\it i.e.\/} impact parameter dependence 
 presented as dependence on inelastic (wounded) participant
number $A$.
Right: energy $\sqrt{s_{\rm NN}}$-dependence. Note that the results
at RHIC (left side and the two highest energies in green on right) 
 apply to central rapidity region $dN/dy$ study, while the AGS/SPS 
energy results are obtained from total  yields and are extrapolated to
the top Brahms result at RHIC.
}
\end{figure}

To study the energy dependence,  we   must assemble  several
different experimental results from different facilities and 
experiments~\cite{edepend}. The results of this extensive 
analysis are shown on right hand side in 
\rf{gammu}. We note that when we are able to consider 
the   total particle abundances, the number of participating nucleons 
is a tacit fit parameter. The lowest energy result is from our 
AGS study~\cite{AGS}, the SPS data we used are from NA49 energy dependence 
exploration at the CERN-SPS~\cite{Gaz,GazPriv}. These results are
for the total particle yields. The highest energy  two results are based on 
studies of RHIC data at 130 and 200 GeV  at central rapidity and address the $dN/dy$
particle yields, with the highest point corresponding to the 
results presented at greatest centrality on the left hand side in \rf{gammu}.

There are several relevant  features in  Fig.\,\ref{gammu}. 
We see on left, that,  for $A>20$, there is no centrality dependence 
in freeze-out 
temperature $T$ and chemical potentials $\mu_{B,S}$ 
(up to the variation  which can be associated with fluctuation
in the data sample).  However, there is a change in values of the
chemical potentials with reaction energy. This is result of rapidly decreasing
baryon density, which due to reduced stopping is distributed over a wider
range of rapidity as the reaction energy increases. 

The middle sections, in   \rf{gammu},
address the phase space occupancies  which were  obtained   
in terms of hadron particle yields.  The quark-side occupancy
parameters could be considerably different, indeed as model
studies show, a factor 2 lower at the discussed 
conditions~\cite{CUP}: the hadron side phase
space size is in general different from the quark-side phase space,
since the particle degeneracies, particle spectra  are quite different. 
We see similar  behavior of  $\gamma_q$ as of $T$ both for volume (that is,
`wounded' participant number $A$)
and energy dependence seen in  Fig.\,\ref{gammu}: the two lowest
energy bins (top AGS and lowest  SPS energy) deviate from the behavior seen 
at all other energies, as do the bins with $A<20$. 
 $\gamma_s/\gamma_q$ as function of centrality 
rises steadily indicating longer lifespan of the fireball with
increasing size. As function of energy,   $\gamma_s/\gamma_q$  reaches 
  a plateau at 30 $A$GeV, with further rise only seen 
in the central rapidity results  at RHIC. We note that, in our analysis, 
there is no saturation of the $\gamma_s$ as we approach the most
central reactions. This is inherent in the data we consider
which includes the yields of $\phi$, and $K^*$. This result is consistent
with the implication that strangeness is not fully saturated in the
QGP source, though it appears over-saturated when measured in 
the hadron phase space. 

 The deviation  at  the most peripheral centrality bins
and at lowest reaction energy 
from trends set by other results could be an indication of 
the change in the reaction mechanism. 
As a threshold of centrality and/or energy are crossed
a relatively small value of $\gamma_q\simeq 0.5$ grows rapidly 
to the maximum allowed by pion condensation condition, 
$\gamma_q\simeq e^{m_\pi/2T}\simeq 1.6$. This behavior signals a transformation
of a chemically under-saturated phase of matter into something novel,
where chemical equilibration is easy, and results in hadronization in
a over-saturated phase of matter.

Independent of the chemical (non-)equilibrium
assumption, the baryochemical potential   $\mu_B=25\pm1$ MeV 
is seen  across  
10 centrality bins. Similarly,  we find strangeness  chemical potential
 $\mu_S=5.5\pm0.5$ MeV (related to   strange quark chemical potential
$\mu_s=\mu_B/3-\mu_S$).  
The most notable  variation,  in  Fig.\,\ref{gammu}, is 
the gradual increase in strangeness phase space 
occupancy $\gamma_s/\gamma_q$  and thus strangeness yield 
with collision centrality.   This effect was predicted and  
originates in an increasing  lifespan of the fireball \cite{impact}.
The over-saturation of the phase space has been also
 expected  due to both, the dynamics of expansion \cite{RHICPred},  
and/or reduction in phase space size as a parton based matter turns
into HG \cite{JRBielefeld}. This latter
effect is also held responsible for the saturation
of light quark phase space $\gamma_q\to e^{m_\pi/2T}$. 
A systematic increase of $\gamma_s$ with collision centrality 
has been reported for several reaction energies  \cite{Kampfer:2003pf}.

\subsection{Hadronization Temperature}\label{Tphase}

Let us now discuss more
in depth the magnitude of the   hadronization temperature which 
at high energy/central collisions we find at $T=140$ MeV.
Some prefer   the statistical hadronization 
to occur at  higher temperature, perhaps as high as$ T=175$ MeV,
a point argued   at this meeting in great detail  
by  Dr. Peter  Braun-M\"unzinger. In his presentation we heard that
he believes  that the lattice results will reach to such high temperature
near to $mu_{\rm B}=0$ and this is where one should expect to see hadronization
We disagree with both claims. For one, 

We note that  {\it chemical equilibrium QCD 
lattice}~\cite{Fodor:2004nz}  results are  mature and yield
  $T =163\pm2$ MeV  when extrapolated
to physical quark mass scale. This result is in a very good agreement
with the prior work on 2 and 3 flavor QCD which brackets this result
by $T_{n_f=2,3}=T\pm 10$ MeV~\cite{Karsch:2001vs}. Moreover,   
 the heavy ion collisions present  a highly 
dynamical environment and one has to pay tribute to this especially
regarding the value of hadronization temperature. 
 For this reason   we do not 
expect to find that the observed hadronization 
condition  $T(\mu_{\rm B})$ will 
line up with the phase boundary curve obtained in 
study of a statistical system in thermodynamic limit on the lattice:

a)   \underline{Dependence on parton collective flow:}\\
  A widely discussed  effect which displaces the 
hadronization condition from the phase boundary  is the expansion dynamics 
of the fireball. When  the collective flow occurs at parton level,
the color charge flow, like a wind,   pushes  out the vacuum~\cite{Csorgo:2002kt},
adding to thermal pressure a  dynamical component.
This  can in general lead to supercooling and a sudden breakup of 
the fireball. We find that this  can 
reduce the effective hadronization temperature by up to 20 MeV~\cite{suddenPRL}.

b) \underline{Dependence on quark chemical equilibration:}\\ 
Lattice result have been discussed  for  2-flavor lattice QCD at
corresponding to   $\gamma_q=1,\gamma_s=0$ (called) and for
2+1 flavor, corresponding to 
$\gamma_q=\gamma_s=1$. While the precise nature of the phase limit
is still under study it appears that  
for 2-flavor case the phase
boundary temperature rises by about 7-10 MeV compared to 
the 2+1 case. We refer to the recent review of lattice QCD 
for further details~\cite{Karsch:2003jg}. 
 Similarly the phase limit in pure gauge case corresponding in
loose sense to 
$\gamma_q=\gamma_s=0$  was seen near or even above $T=200$ MeV. 

These results do suggest that presence and the number of quarks matters
regarding  the precise location of the phase boundary and its   nature.
 Its  importance
could be greatly enhanced, should  the over-saturation of quark phase space 
have the same effect as  would additional quark degrees of freedom. These are
known to   cause even for $\mu_{\rm B}=0$ 
the conversion of the phase crossover into a  1st-order phase transition
which would, with these additional degrees of freedom, be 
expected at just the temperature we find in the SHM analysis.

We can be nearly sure that the chemical conditions matter and can
displace the transition temperature. Because the degree of 
equilibration in the QGP depends on the collision energy, as does 
the collective expansion velocity, we cannot at all expect a 
simple hadronization scheme appropriate for the hadronization
of nearly adiabatically expanding Universe. 

Leaving this issue we note that, in the data analysis assuming
chemical equilibrium, we  find  $T=155\pm8$ MeV for the 
 chemical equilibrium and strangeness non-equilibrium  
freeze-out, see \rf{gammu}. The error is our estimate of the propagation 
of the systematic data error, combined with the fit uncertainty;
the reader should note that the   error comparing centrality
to centrality is negligible. The freeze-out temperature is for the 
semi-equilibrium and equilibrium model about  10\% greater than 
the full chemical non-equilibrium freeze-out. 

  This result for $T$, in the equilibrium case, is in mild
disagreement (1.5 s.d.) with earlier  equilibrium 
fits  \cite{BDMRHIC,Broniowski:2003ax}. 
This, we believe, is due  to some
  differences in data sample used, specifically,
the  hadron resonance production results used
provide  a very strong constraint for the fitted 
temperature, and more complete 
treatment by  SHARE of  hadron mass spectrum.

Interestingly, it seems that the general consensus about the  chemical
equilibrium best analysis result
is in gross disagreement with the   results advanced  
at this meeting by  Dr. Peter  Braun-M\"unzinger.

\subsection{Physical Properties of the Fireball}\label{physres}

\begin{figure}[!bt]
 \hspace*{-.6cm}\psfig{width=8.5cm,figure=\pathnow 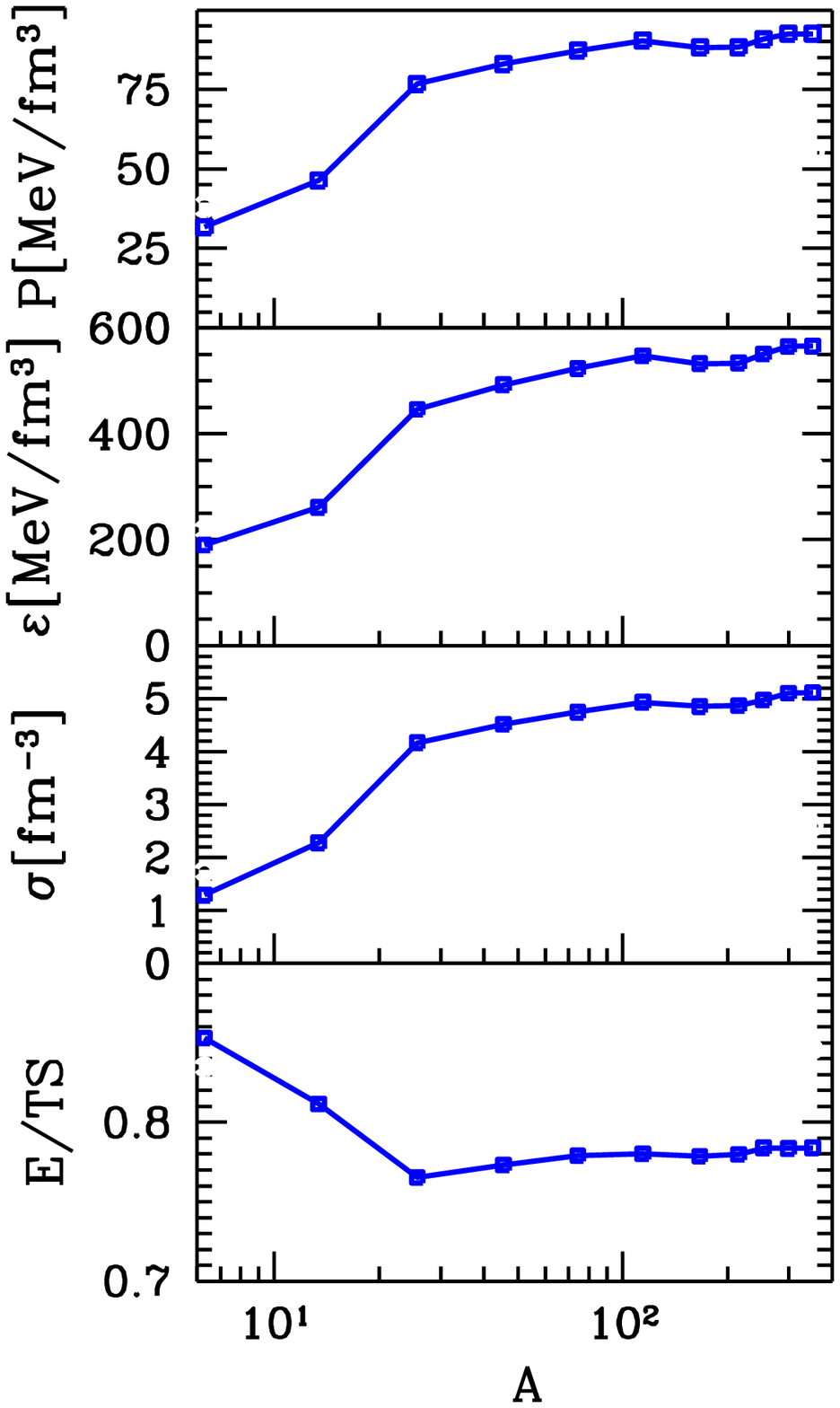}
\hspace*{-.5cm}\psfig{width=8.5cm,angle=0,figure=\pathnow  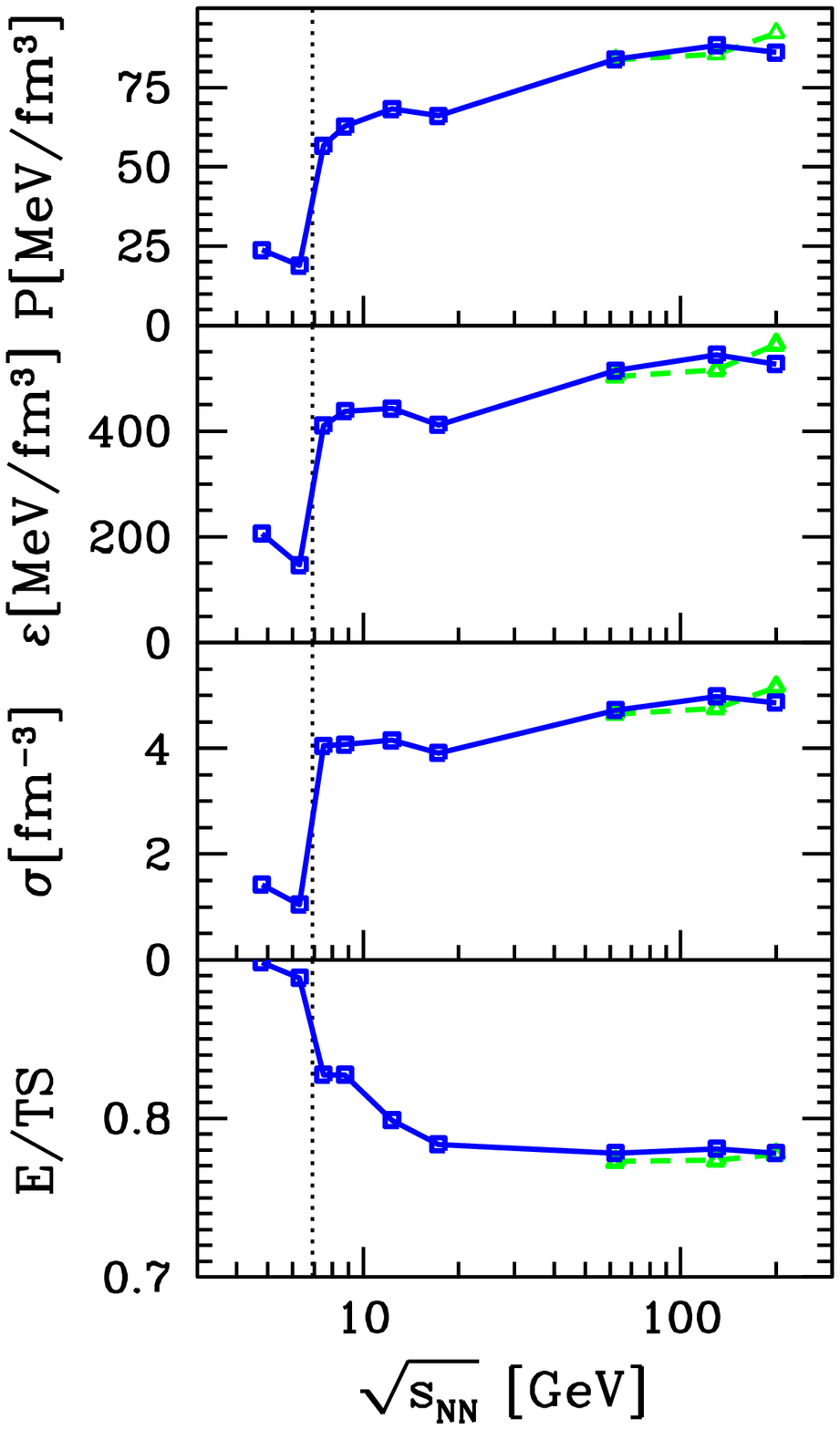}
\vspace*{-0.6cm}
\caption{\label{Phys}
 From top to bottom:  pressure $P$, energy density $\epsilon=E/V$, 
entropy density $S/V$ and   $E/TS$; Left:
as a function of centrality, Right as function of reaction energy.
The results on left are for central rapidity, and on   right are for the full
phase space coverage, except for the high energy RHIC results (open triangles, green)
which are for central rapidity.  Open circles on left apply to the 
semi-non-equilibrium analysis with $\gamma_q=1, \gamma_s\ne 1$, all other 
results are for the chemical  nonequilibrium analysis.
  }
\end{figure}

We now turn our attention to the physical properties of the hadronizing 
fireball obtained summing individual contributions made 
by   made of each of the hadronic particles  
 produced. Often particles observed experimentally
dominate ({\it e.g.\/} pions dominate pressure, kaon strangeness yield etc).
However, it is important to include in this yields  of
particles  predicted in terms
of  the SHM fit to the available results. Again, on left in Fig.\,\ref{Phys}.
we show the behavior as function of impact parameter and on right as function
of energy. From top to bottom we show the pressure $P$, energy density $\epsilon$, 
entropy density $\sigma$, and the dimensionless  ratio $\epsilon/T\sigma=E/TS$.
All  contributions are evaluated using relativistic expression, see \cite{CUP}.

When we fitted   particle rapidity  yields,  the global 
fitted yield normalization factor is $dV(A)/dy $, for total
particle yields it is $V$ which is also a function of centrality trigger.
When we consider ratios of two 
bulk properties e.g.  $E/TS$,  the results are in general smoother
indicating cancellation in the error propagating from the fit. 
The overall error is of the magnitude of the particle yield, for
example pressure is dominated by pions and hence its precision 
is limited by this error. However, on left, the point-to-point error
is minimal as the systematic error is common. On right, the absolute error matters
as the fluctuations in the results presented show.  
 
We note, in \rf{Phys}, that 
as the reaction energy passes the volume threshold  $A=20$ and even more so, 
the energy threshold $6.26\,{\rm GeV}<\sqrt{s_{\rm NN}}<7.61\,{\rm GeV}$
the hadronizing fireball becomes much denser.
The   entropy density jumps by factor 3--4, 
and the energy and baryon number density by a factor 2--3.
The hadron pressure jumps up from $P=25\, {\rm MeV}/{\rm fm}^3$ initially
by factor 2 and ultimately more than factor 3. There is 
a gradual increase of $P/\epsilon=0.115$ at low reaction energy 
to 0.165 at the top available energy.  
It is important  to note that exactly the same behavior of the 
fireball physical properties   arises both as function of 
reaction volume   and reaction energy. This is the case   for both
the physical properties and the
statistical parameters. We believe that this shows a common change in
the physical state created as function of energy available and reaction
volume.

\section{Search for an  Energy Threshold}\label{compare}
\subsection{Kaon to pion ratio}\label{Kpisec}
One of the most interesting questions is if there is 
an  energy threshold for the formation of a new state of matter. 
An important observable of the deconfined phase of matter is 
aside of strangeness, the high entropy content, which is
arising from broken color bonds. The observable for both is 
the K$^+/\pi^+ \propto \bar s/\bar d$ yield ratio~\cite{Glendenning:1984ta}. This 
ratio has been studied experimentally~\cite{Gaz}  and a pronounced
 horn structure  arises. We can describe this structure in 
our study of the particle yields only within the chemical non-equilibrium 
model. Although this change is  associated a rather sudden modification of 
 chemical conditions in the dense
matter fireball, this effect is caused   by two distinct phenomena: the 
rapid rise in strangeness $\bar s$ production below,  and  a   rise in 
the antiquark $\bar d$  yield above a energy reaction threshold. 

\begin{figure}[ht]
\vskip-.4cm
\parbox[t]{8.2cm}{\psfig{width=8.2cm,height=8.8cm,figure=\pathnow  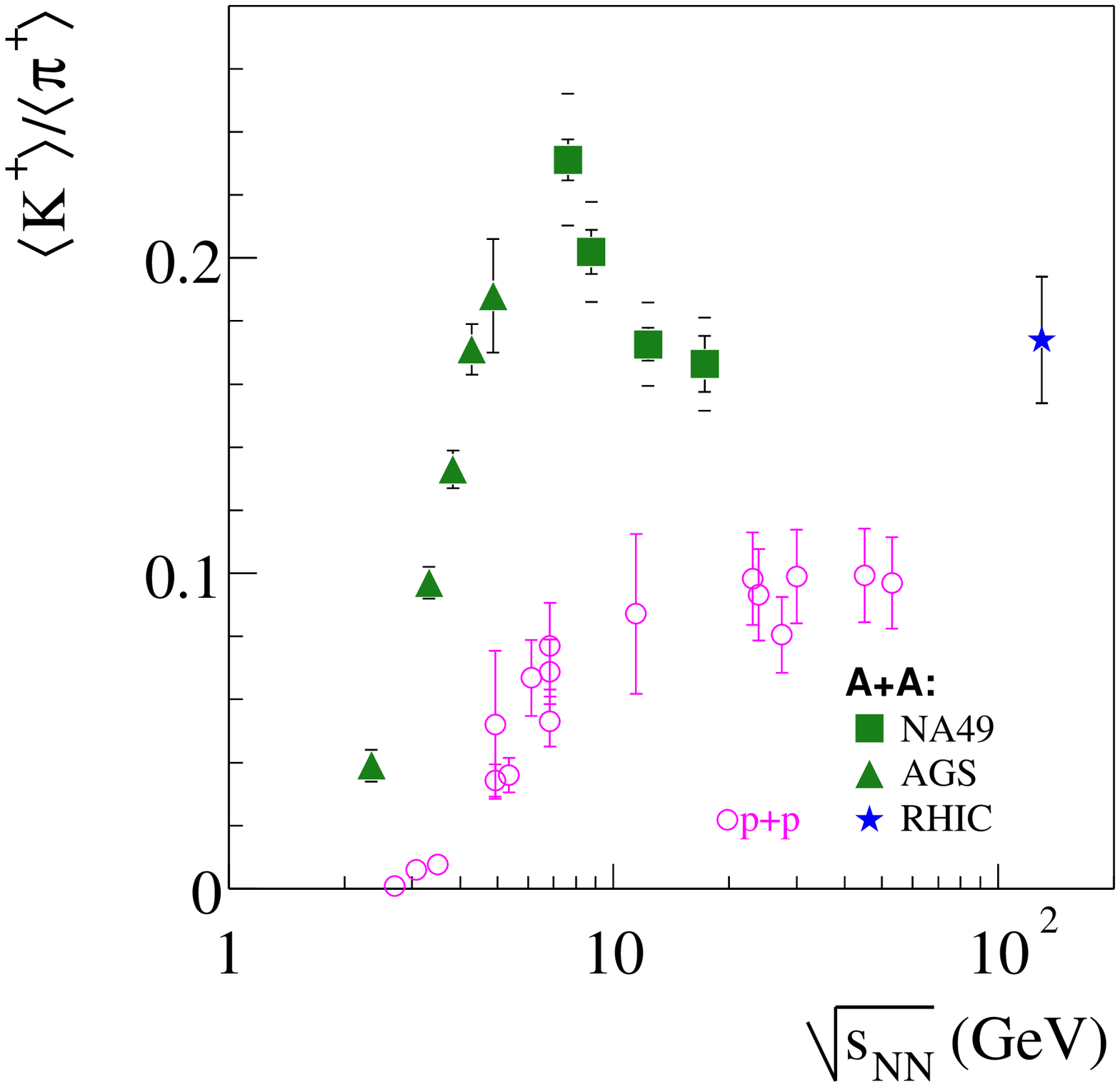}}
 \vskip-.25cm
\hspace*{1.1cm}\psfig{width=7.77cm,height=6.9cm, figure=\pathnow 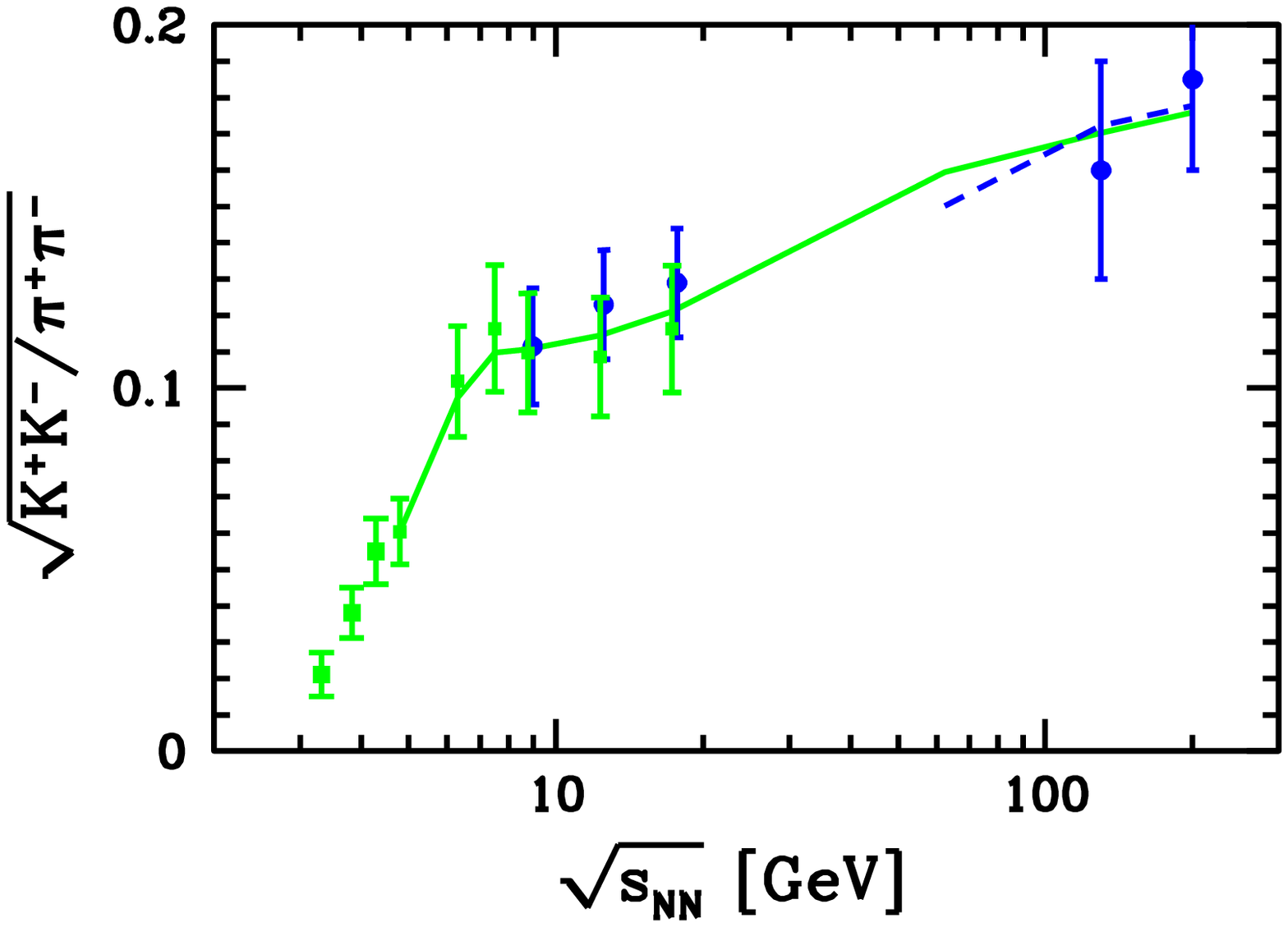}
\vskip-14.75cm\hspace*{8.2cm}
\parbox[t]{8.2cm}{\psfig{width=7.8cm,height=4.1cm,
                                    figure=\pathnow  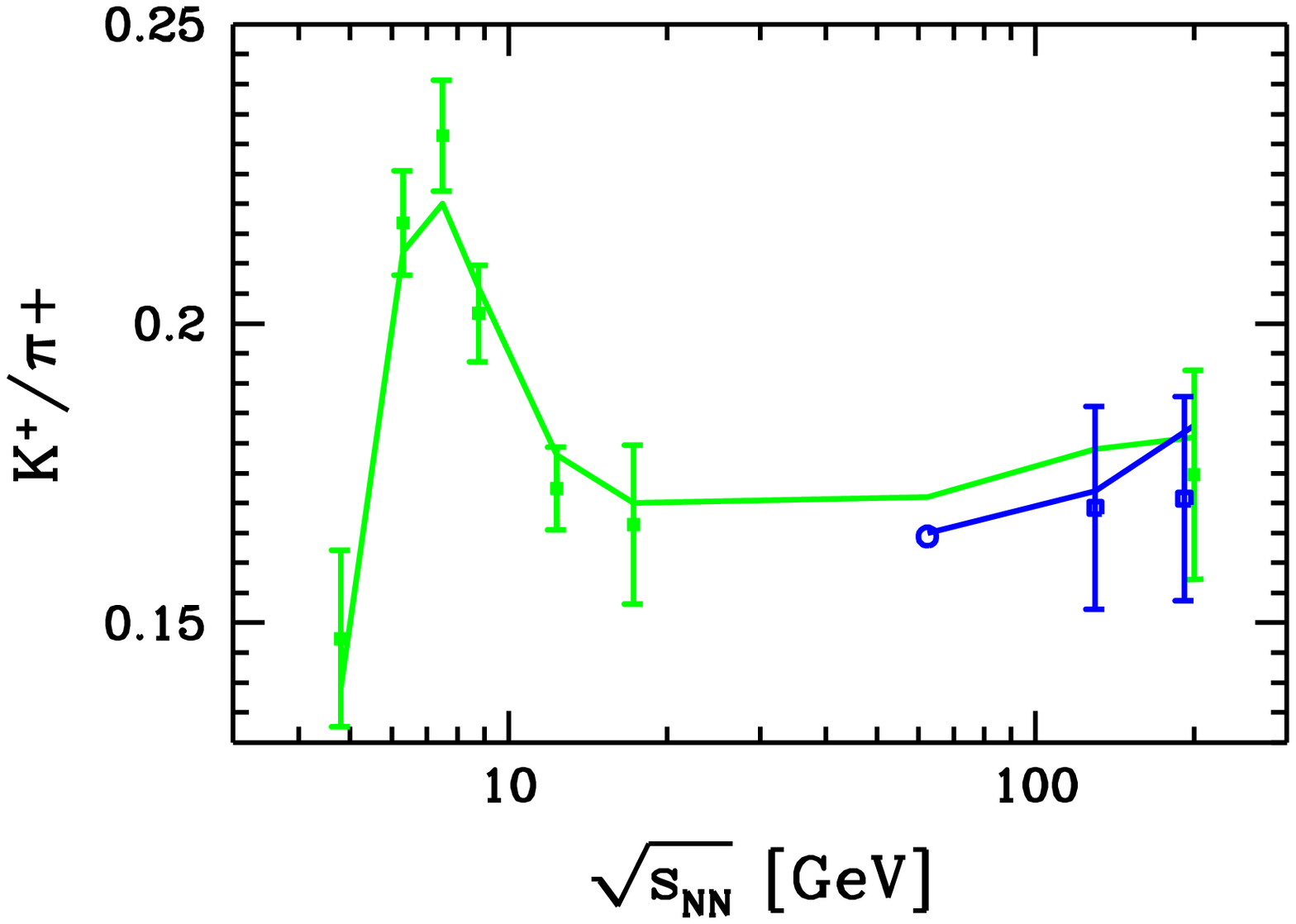}}
\vskip 4.0cm
\hspace*{8.2cm}
\parbox[t]{8.2cm}{\psfig{width=7.77cm,height=3.36cm, figure=\pathnow  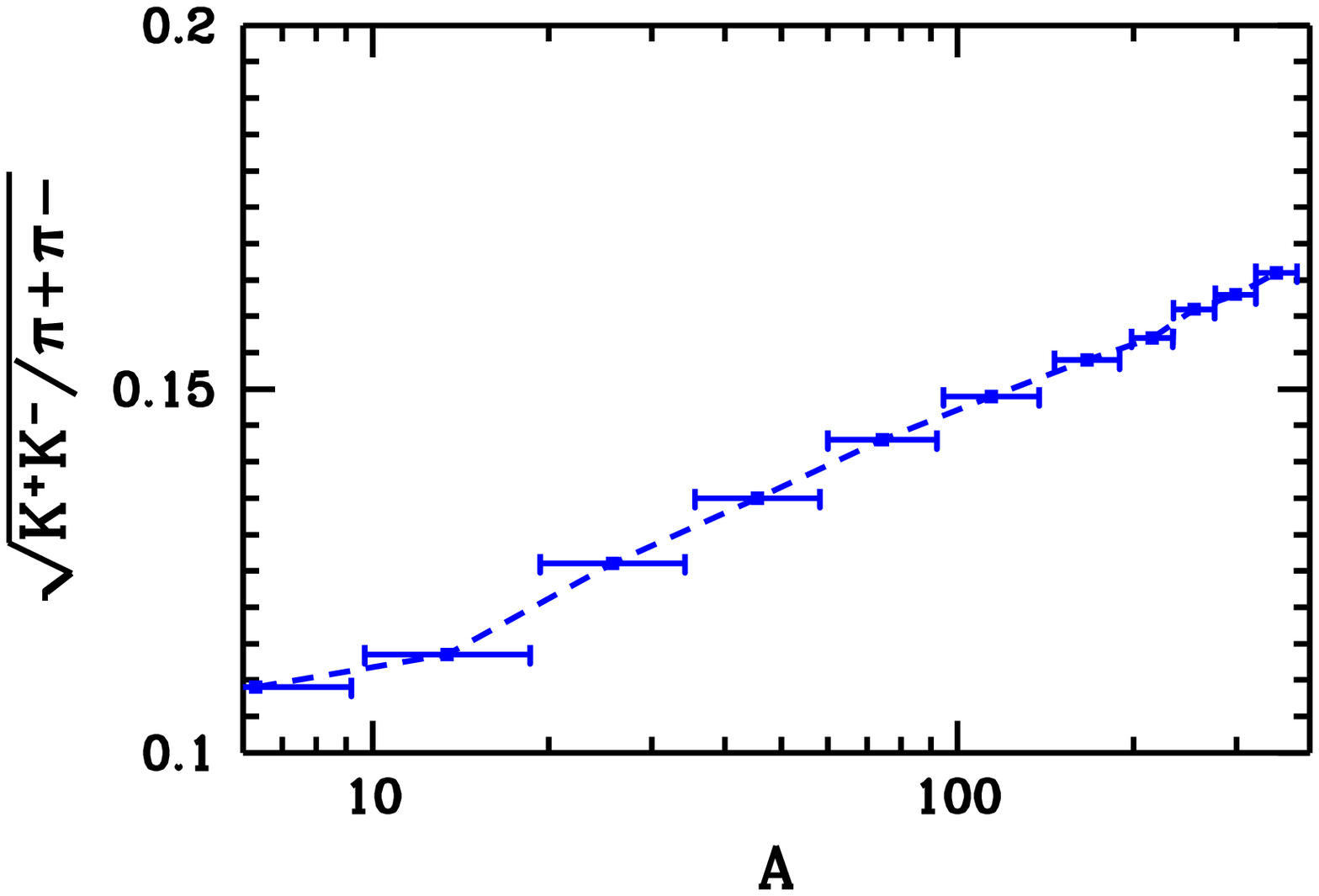}}
\vskip 2.6cm
 \caption{\label{Kpisqrt}
Top: Total yield-ratio of yields of  $K^+/\pi^+$
for nuclear (filled symbols, green) and 
elementary interactions (open symbols, violet);
(courtesy of NA49 collaboration~\protect\cite{Lee03}).
On right: a fit to this $K^+/\pi^+$ ratio using SHM, at high energy also
for central rapidity RHIC, line guides the eye between the fitted values.
Bottom: the  
$K/\pi$ ratio for nuclear  collisions, 
squares on left (green): full phase space on left, circles (blue)
 are central rapidity results, left
 as function of collision energy and right as function of 
participant number at $\sqrt{s_{\rm NN}}=200$ GeV.
}
\end{figure}

The measured ${\rm K}^+/{\pi}^+$ ratio by NA49
 is shown at top left of 
 \rf{Kpisqrt}, where for comparison we also show the $pp$ results. On
right top, we present our results reduced to the  correct relative scale, 
both for the total yield ratio  for
the AGS--RHIC energy range, and for
the central rapidity results from RHIC. 
The solid line guides the eye to the fit
results we obtained.   
To show that the  $K^+/\pi^+$ ratio  drop  is due  to a decrease in baryon  density
which leads to rise in the  $\bar d$ yield,
 we show in bottom section of  \rf{Kpisqrt},
 the nearly  baryon density  independent  ${K}/{\pi}$ double ratio ratio \req{Kdpi}, 
on left as function of  $\sqrt{s_{\rm NN}}$,
and on right as function of centrality of the reaction for 
 $\sqrt{s_{\rm NN}}=200$ GeV:
\beql{Kdpi}
\frac{K}{\pi}=\sqrt{  \frac{K^+}{\pi^+}\, \frac{K^-}{\pi^-}}.
\eeq
Both upper and lower portion of  \rf{Kpisqrt} are also drawn on same relative scale.

Seen how the horn specifically arises in the one  $K^+/\pi^+$,
one can wonder if this is really a physical effect and how, in qualitative terms, 
 a parameter $\gamma_q$, which controls the light
quark yield, can help explain   the horn structure seen in top of  \rf{KPi}. 
We observe that this   horn structure  in the ${\rm K}^+/\pi^+$ ratio  
traces out  the final state  valance quark ratio $\bar s/\bar d$, and in language of
quark phase space occupancies $\gamma_i$ and fugacities $\lambda_i$, we have: 
\begin{equation}\label{KPi}
{{\rm K}^+\over \pi^+}\to {\bar s\over \bar d} \propto F(T)
  \left({\lambda_s\over \lambda_d}\right)^{\!-1} {\gamma_s\over \gamma_d}
  =
   F(T)\left(\sqrt{\lambda_{I3}}{\lambda_s\over \lambda_q}\right)^{\!-1} 
                                   {\gamma_s\over \gamma_q}.
\end{equation}

In chemical equilibrium models $\gamma_s/\gamma_q=1$, 
and the  ${\rm K}^+/\pi^+$ ratio   and its horn 
must arise solely from the variation in the ratio $\lambda_s/ \lambda_q$ and the 
change in temperature $T$ which both are usually smooth function of reaction 
energy. The isospin factor $\lambda_{I3}$ is insignificant in
this consideration.

As collision energy is increased, increased hadron yield leads to 
  a decreasing $\lambda_q=e^{\mu_{\rm B}/3T}$.
We recall the smooth decrease of $\mu_{\rm B}$ with 
reaction energy   seen in bottom panel in \rf{gammu}.
The two chemical fugacities  $\lambda_s$ and $\lambda_q$ are
coupled by the condition that the strangeness is conserved. This 
leads to a smooth  $\lambda_s/ \lambda_q$. The chemical
potential effect is suggesting a smooth increase in the K$^+/\pi^+$ ratio. 
 
 For the interesting 
range of freeze-out temperature, $F(T)$ is a smooth  function of $T$. Normally,
one expects that $T$ increases with collision energy, hence on this ground 
as well we expect an
monotonic  increase  in the ${\rm K}^+/\pi^+$ ratio as function of reaction
energy.  
With considerable effort, one can arrange the chemical equilibrium fits to 
bend over to a flat behavior at  $\sqrt{s_{\rm NN}^{\rm cr}}$.   
It is quasi impossible to generate the horn 
with chemical equilibrium model. To accomplish this, an additional parameter
appears necessary, capable to change rapidly when hadronization conditions
change. This is $\gamma_q$. It presence also allows $T$ to vary in non-monotonic
fashion, as is seen in \rf{gammu}.

\subsection{QGP degrees of freedom and $s/S$ ratio}\label{sSrat}
The full SHM is capable to describe the data, and 
we now show that it can pinpoint the properties of the 
phase of matter that was created early on in the reaction. To see
this we consider  what we learn from the final state data about
strangeness and entropy 
production. For this purpose we consider both the 
specific per baryon and per entropy yield of strangeness. In addition, 
we look at the cost in thermal energy to make strangeness. All these
quantities are  nearly independent of the  dynamics of hadronization,
since they are related to processes occurring early on, `deep'
 inside the collision region, and long before hadronization. 

In the  QGP, the dominant entropy production 
occurs during  the initial glue thermalization, and the  thermal
strangeness production occurs in parallel and/or just a short time later~\cite{Alam:1994sc}.  
The entropy  production occurs predominantly 
early on in the collision during  the thermalization phase.  
Strangeness production by gluon fusion 
is  most effective in the early, high temperature environment,
however it continues to act during the evolution of
the   hot deconfined phase until hadronization \cite{RM82}.
Both strangeness and entropy are nearly conserved in
the evolution towards hadronization
and thus the final state hadronic yield analysis value for $s/S$ is closely
related to the thermal processes  in the fireball at $\tau\simeq 1$--2 fm/c. 
We believe that for reactions in which the system approaches strangeness
equilibrium in the QGP phase, one can expect a prescribed ratio of 
strangeness per entropy, the value is basically the ratio 
of the QGP degrees of freedom.

We estimate the magnitude of $s/S$ deep in the QGP phase, considering  
the hot   stage of the reaction. For  an   equilibrated 
non-interacting QGP phase with perturbative properties:
\beql{sdivS}
{s \over S}\equiv\frac{\rho_{\rm s}}{\sigma}     =
\frac{ (3/\pi^2) T^3 (m_{  s}/T)^2K_2(m_{  s}/T)}
  {(32\pi^2/ 45)  T^3 
    +n_{\rm f}[(7\pi^2/ 15) T^3 + \mu_q^2T]}=
{0.027\over {1+ 0.054 (\ln \lambda_q)^2} }\,.
\eeq
Here, we used for the number of flavors $n_{\rm f}=2.5$ and $m_{  s}/T=1$. We 
see that the result is a slowly changing function  of $\lambda_q$;
for large $\lambda_q\simeq 4$, we find at modest SPS energies, the 
value of $s/S$ is reduced by 10\%. Considering 
the slow dependence on $x=m_{  s}/T\simeq 1$ of $W(x)=x^2 K_2(x)$ there is 
minor dependence on the much less variable temperature $T$.

The dependence on the degree of chemical equilibration 
which dominates is easily obtained separating the different 
degrees of freedom:
\beql{sdivS2}
{s \over S}=0.027 {\gamma_s^{\rm QGP} \over 
  {0.38 \gamma_{\rm G}^{\rm QGP}+ 
         0.12 \gamma_s^{\rm QGP}+
         0.5\gamma_q^{\rm QGP} + 
         0.054 \gamma_q^{\rm QGP} (\ln \lambda_q)^2}}\,.
\eeq
We assume  that the  interaction effects are at this level of the 
discussion canceling.  Seen \req{sdivS2} we expect to see a gradual 
increase in $s/S$ as the QGP source of 
particles approaches chemical equilibrium with increasing 
collision energy   and/or  increasing volume. 

We repeat that it is important to keep in mind 
that this ratio $s/S$ is established early on in the 
reaction,    the above relations, and the associated 
chemical conditions  we considered, apply to
 the early hot  phase of the fireball.
 At hadron freeze-out the 
QGP picture used above does not apply. Gluons are likely
to freeze faster than quarks and both are subject to much
more complex non-perturbative behavior. However, the value of
$s/S$ is nearly preserved from the hot QGP to the  final state
of the reaction.
 
How does this simple prediction compare to experiment?
Given the statistical parameters, we can evaluate the 
yields of particles not yet measured and obtain
the rapidity yields of  entropy, net baryon number, net strangeness,
and thermal energy, both for the total reaction
system and also  for the central rapidity condition, also
 as function of  centrality. In passing, we note
that,   in the most central reaction bin at RHIC-200,
  $dB/dy\simeq 15$ baryons per unit rapidity interval, implying a rather 
large baryon stopping in the central rapidity domain. 

\begin{figure}[!bt]
\psfig{width=8.5cm,angle=0,figure=\pathnow  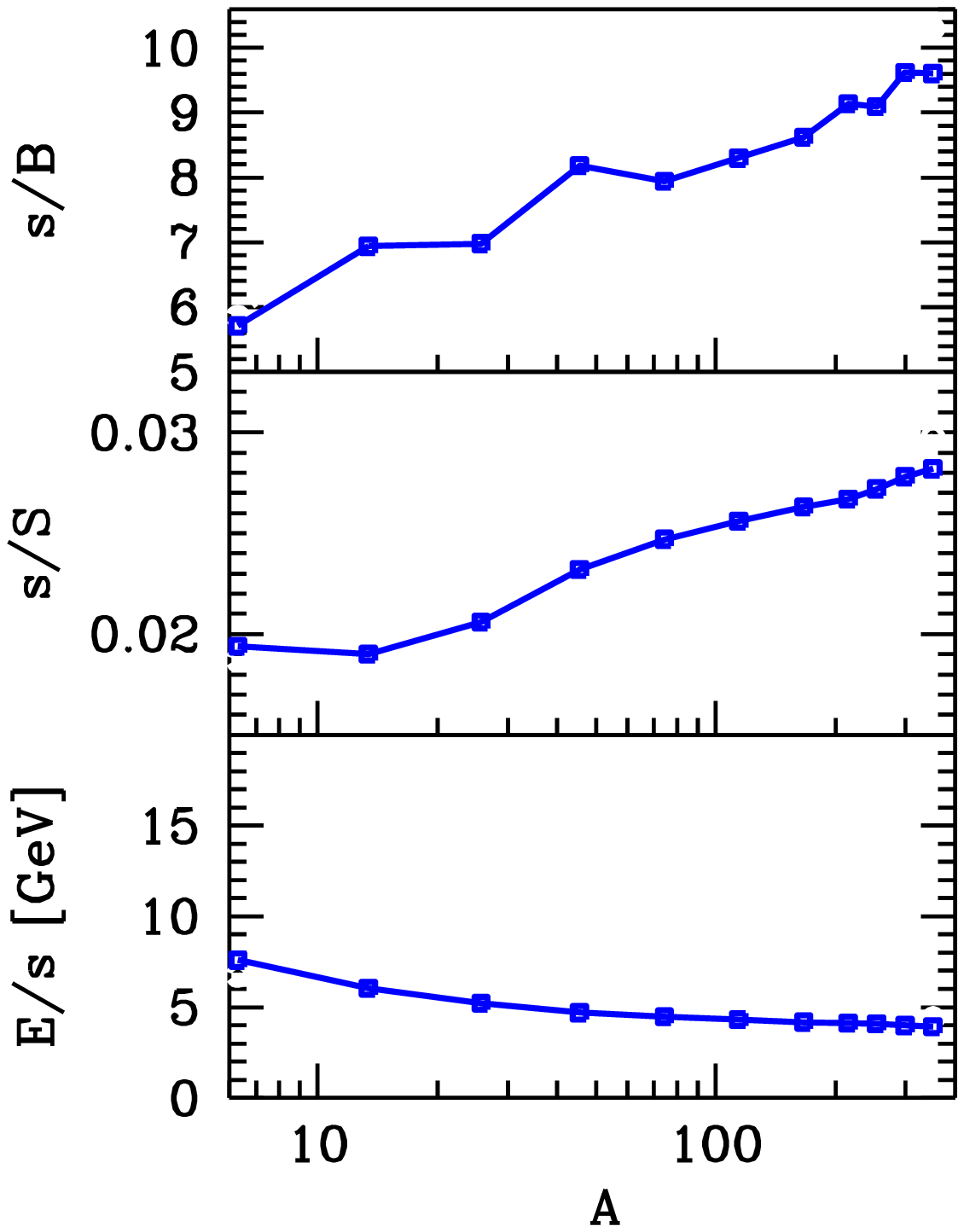}
\hspace*{-.6cm}\psfig{width=8.5cm,figure=\pathnow 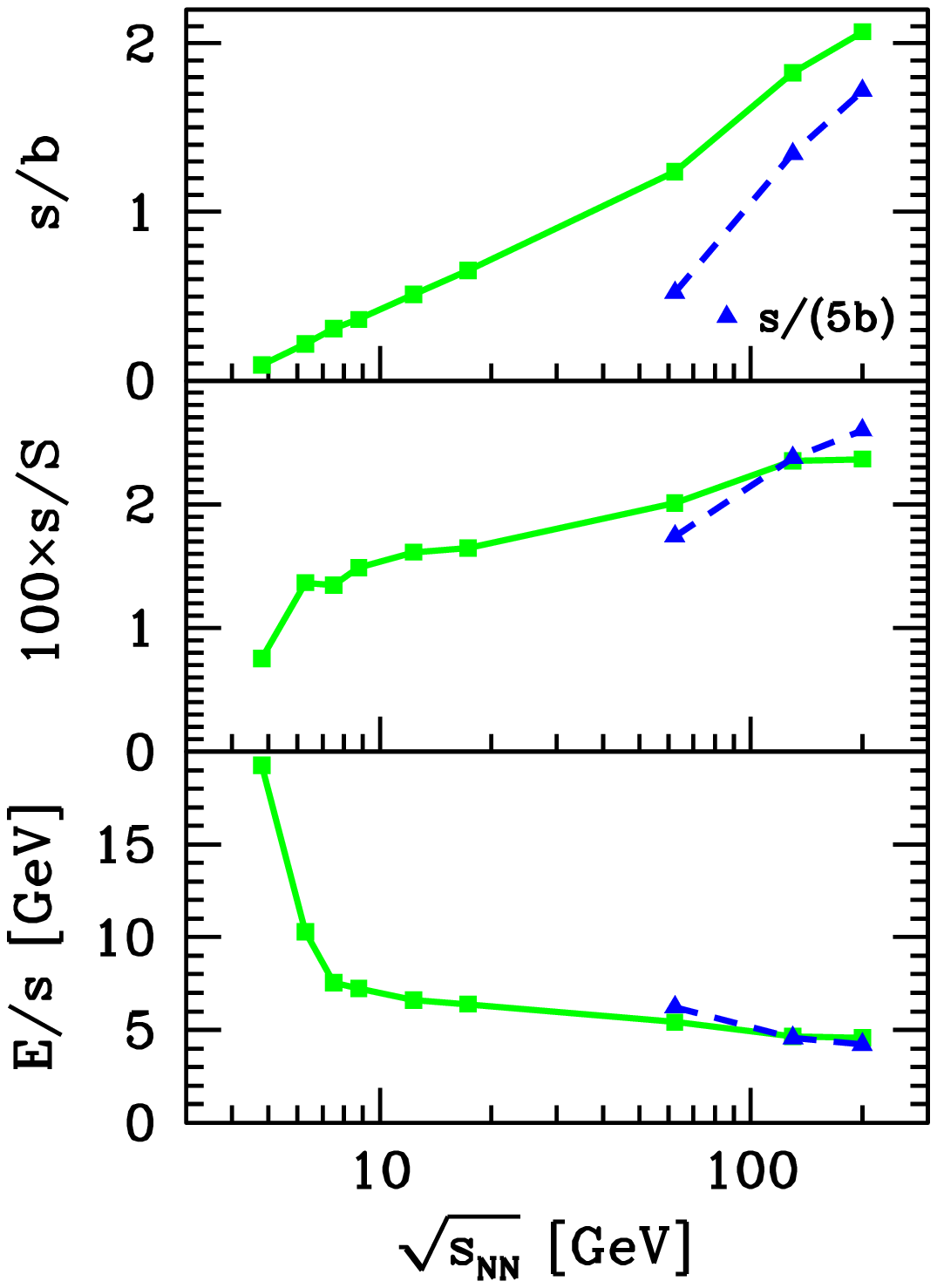}
\vspace*{-0.6cm}
\caption{\label{PEST}
Strangeness per net baryon $s/B$, 
strangeness per entropy $s/S$,   and   $E_{\rm th}/s$ the thermal energy  
cost to make strangeness.  Left:
as a function of centrality, Right as function of reaction energy.
The results on right 
include the central rapidity conditions at RHIC energies
(dashed, blue) lines. The actual results are the symbols, the lines guide the 
eye.  
  }
\end{figure}

The rise of strangeness yield with 
centrality is faster than the rise of baryon number yield:
 $(ds/dy)/ (dB/dy)\equiv s/B$ is seen in  the top left panel 
 in Fig.\,\ref{PEST}. The solid (blue) lines are for the chemical
nonequilibrium central rapidity yields of particles at RHIC-200. 
Solid (green) lines, on right,
 are for total hadron yields and thus total yields of the considered 
quantities, {\it e.g.\/} strangeness, entropy.
For the  most central head-on reactions, we  reach   
at RHIC-200 $ s/B =9.6\pm 1$.   
  
In the middle  panel  in \rf{PEST}   we compare strangeness with 
entropy production $s/S$, which we just evaluated theoretically. Again,
 on left, as function of participant number $A$  at RHIC-200
 and, on right, for the total reaction system 
for the most central 5-7\% reactions. On left, we see a 
smooth transition from a flat peripheral 
behavior where $s/S\lessim 0.02  $  to smoothly increasing $s/S$
 reaching $s/S\simeq 0.028$
in most central reactions. This indicates that even at  RHIC-200
for the more central reactions some   new mechanisms of
strangeness production becomes activated. On right, we see that 
the change on $s/S$ is much more drastic as function of reaction
energy. After an initial rapid rise the further increase occurs beyond the
threshold energy at slower pace. 

In the bottom panel, on left, we see the thermal energy cost $E_{\rm th}/s$ 
of producing a pair of strange quarks as function of the size of the participating 
volume ({\it i.e.\/} $A$) This quantity shows a smooth 
drop which can be associated with transfer of thermal energy into collective
transverse expansion after strangeness is produced. Thus, it seems that the 
cost of strangeness production is independent of reaction centrality. The
result  is different when we consider $\sqrt{s_{\rm NN}}$ dependence of this 
quantity, see bottom panel on right. There is a very  clear change in the 
energy efficiency of making strangeness 
at the threshold energy. We will return to discuss possible reaction mechanisms below.

\section{Final remarks}\label{final}
\subsection{Phase boundary and hadronization conditions}\label{PhaseB}
The  chemical freeze-out  conditions we have determined presents,    in the 
$T$--$\mu_{\rm B}$ plane, a more complex picture than naively expected,
 see     \rf{Tmu}. Considering results
 shown in   \rf{gammu}, we are able to assign to 
each point in the $T$--$\mu_{\rm B}$ plane the associate value of 
 $\sqrt{s_{\rm NN}}$. The RHIC $dN/dy$ results are to outer left.
They are followed by RHIC and SPS $N_{4\pi}$ results. The dip
corresponds to the 30 and 40 $A$GeV SPS results. The top right is 
the lowest  20 $A$GeV SPS and top 11.6 $A$GeV AGS energy range. 
To guide the eye, we have
added two lines connecting the fit results.  We see that the  
chemical freeze-out temperature 
$T$ rises for the two lowest reaction energies  11.6 and 20 $A$ GeV
 to near the Hagedorn temperature,  $T=160$ MeV, of boiling hadron  
matter. 

\begin{figure}[!tb]
\centerline{\psfig{width=10cm,figure=\pathnow     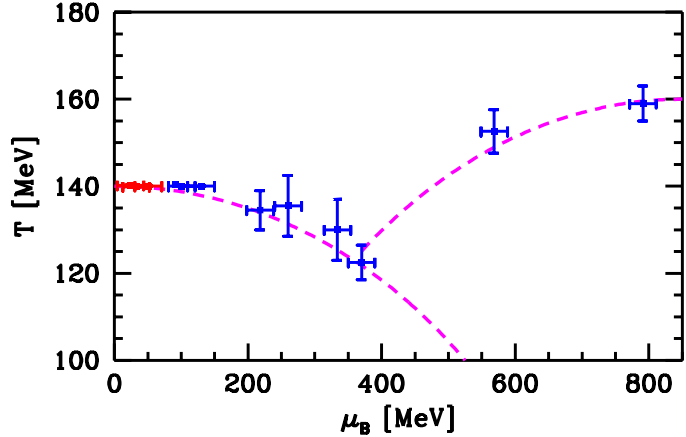}}
\caption{\label{Tmu}
$T$--$\mu_{\rm B}$ plane with points obtained in the SHM fit.  
}
\end{figure}

Once the chemical non-equilibrium is allowed for, the data fit turns
to be much more precise, and the 
picture of the phase boundary with smaller `measurement' error
reveal a much more complex structure, and contains physics
details prior analysis based on a rudimentary model could not uncover. 
The shape of the  hadronization boundary,  shown in \rf{Tmu} 
in the $T$--$\mu_{\rm B}$ plane,   is the result
of a complex interplay between the dynamics of 
heavy ion reaction, and the properties of both phases of
matter, the inside of the fireball,  and the hadron phase we
observe.  The dynamical effect,  
capable to shift the location in temperature of 
the expected phase boundary is due to 
the expansion dynamics of the fireball which occurs at parton level,
and effects of chemical nonequilibrium, see subsection \ref{Tphase}
for full discussion. 
 
Possibly, not only the location,
but also the {\it nature} of the phase boundary can be  modified by 
variation  of $\gamma_i$. We recall that for the 2+1 flavor case,
there is a critical point at finite baryochemical potential
with $\mu_{\rm B}\simeq 350$ MeV~\cite{Fodor:2004nz}. 
 For the case of 3 massless flavors, 
there can be  a 1st order transition 
at all $\mu_{\rm B}$ \cite{Karsch:2001vs,Peikert:1998jz,Bernard:2004je}. 
Considering a classical particle system, one easily sees 
that  an  over-saturated
 phase space, {\it e.g.\/}, with  $\gamma_q=1.6,\ \gamma_s\ge \gamma_q $  
for the purpose of the study of the phase transition acts   
 as being equivalent to a system with 
3.2 light quarks and 1.6 massive (strange)
quarks present in the confined hadron phase. 

 Considering 
the sudden nature of the fireball breakup 
seen in several observables~\cite{RBRC}, 
we conjecture that the hadronizing fireball leading to
 $\gamma_s>\gamma_q=1.6$  super-cools and experiences  
 a true 1st order  phase boundary   corresponding also  at small 
$\mu_{\rm B}$. The system we observe 
in the final state prior to hadronization is mainly a quark--antiquark
system with gluons frozen in prior expansion cooling of 
the QCD deconfined parton fluid. The quark dominance is necessary
to understand {\it e.g.\/}  how  the azimuthal asymmetry $v_2$  
varies for different particles~\cite{Huang:2005nd}.

These quarks and antiquarks have, in principle,  at that stage a significant 
 thermal mass. The evidence for this is seen  
in \rf{Phys} in its two bottom panels, showing the
 dimensionless variable $E/TS$. The energy  end entropy  per particle
 of  non-relativistic and semi-relativistic classical 
particle gas comprising both quarks and antiquarks 
is  (see section 10, \cite{CUP}):
\begin{equation}
{ E\over N}\simeq m+3/2\, T+\ldots,\quad
{ S\over N} \simeq 5/2+m/T+\ldots,\qquad   
{E\over TS}\simeq {m/T+3/2\over m/T+5/2}\ .
\end{equation}
 It is thus possible to interpret the fitted value $E/TS\to 0.78$ 
 in terms of a quark matter made of particles with 
  $m\propto aT$, $a=2$  which is close
to what is  expected based on 
thermal QCD~\cite{Petreczky:2001yp}.  

\subsection{Looking forward to LHC} \label{LHC}
We expect considerably more violent transverse expansion of the 
fireball of matter created at LHC.  The kinetic energy of this
transverse motion must be taken from the thermal energy of the expanding matter,
and ultimately this leads to local cooling and thus a reduction in the number of 
quarks and gluons. The local entropy density decreases, but the expansion
assures that the total entropy still increases. Primarily, gluons are
feeding the expansion dynamics, while strange quark pair yield, being  weaker
coupled remain least influenced by this dynamics. Model calculations 
show that this expansion yields an increase in the final QGP 
phase strangeness occupancy $\gamma_s$  \cite{Raf99R}. 
This mechanism, along with the required depletion of the non-strange degrees
of freedom, in the feeding of the expansion,   assures an increase in the 
$K/\pi$ ratio given the nearly 30-fold increase of collision energy. 

Depending on what we believe to be a valid hadronization temperature for
a fast transversely expanding fireball, the possible maximal enhancement in the 
$K/\pi$ ratio may be in the range of a factor 2--3. 
Perhaps even more interesting than the $K/\pi$ ratio enhancement would be the 
enhancement anomaly in strange (antibaryon) yields. With $\gamma_s\gg 1$, we  
 find that the more  strange  baryons and antibaryons are more
abundant than the more `normal' species.   Specifically of interest would 
be $(\Omega^-+\overline\Omega^+)/(h^++h^-)$, 
$(\Xi^-+\overline\Xi^+)/(h^++h^-)$, and $2\phi/(h^++h^-)$ which should 
show a major, up to an  order of magnitude shift in relative production
strength.  Detailed predictions for the yields of these particles
require considerable extrapolation of physics conditions from the RHIC to LHC
domain and this work is in progress~\cite{LHCPred}.

Ultimately,   strange $s$, and $\bar s$ quarks can exceed in abundance the 
light quark component, in which case we would need to rethink
in much more detail the distribution of global particle yield.
The ratios of neutral and charged hadrons would undergo serious change.

 Regarding charm, we note that situation will not become 
as severe. Given the large charm quark mass, we  expect that
most of charm quark yield is due to first hard interactions of
primary partons. For this reason, the yield of strange and light 
quarks, at time of hadronization, exceeds by about a factor 100  or more 
that of charm at central rapidity. Thus, even though charm 
phase space occupancy at hadronization may largely exceed the 
chemical equilibrium value, seen the  
low hadronization temperature, {\it e.g.\/}, $m_c/T\simeq 10$, it 
takes a factor $\gamma_c\simeq e^{10}/ 10^{1.5}=700$ to compensate 
hadron yield suppression due to the high charm mass. Said differently,
while strange quarks can compete in abundance  with light quarks 
for $m_s/T\simeq 1$, charm (and heavier) flavor(s) will remain suppressed,
in absolute yield, at the temperatures we can make presently in
 laboratory experiments. 

\subsection{Highlights}
We have shown that strangeness, and entropy,
at   SPS and  RHIC are  well developed  tools
allowing  the detailed  study of hot  QGP phase. Our 
detailed discussion of hadronization analysis results
 has further shown that a systematic study 
of strange hadrons  fingerprints the  properties 
of  a new state of matter at point of its breakup into final
state hadrons. 

We have shown that it is possible to describe the `horn' in 
the $K^+/\pi^+$ hadron ratio within the chemical non-equilibrium 
statistical hadronization model. We have shown that  appearance
of this structure is related to a rapid change in the properties 
of the hadronizing matter.  
Of most theoretical relevance and interest are the 
implications of non-equilibrium hadronization on the possible
change in the location and {\it nature} of the 
 phase boundary.

In summary, we have presented  interpretation of the
experimental soft hadron data production and discussed
the production of strangeness and entropy that this 
analysis infers. We have seen, in quantitative way, how  
the relative strangeness and entropy production
in most central high energy heavy ion collisions agrees with 
quark-gluon degree of freedom counting in hot primordial matter
where the values of these quantities have been established.  

\subsection*{Acknowledgments}
Work supported in part by a grant from: the U.S. Department of
Energy  DE-FG02-04ER4131. 
LPTHE, Univ.\,Paris 6 et 7 is: Unit\'e mixte de Recherche du CNRS, UMR7589.
JR thanks Bikash Sinha and Jan-e Alam and the  organizers of  the 5th
 International Conference on Physics and Astrophysics of Quark Gluon Plasma,
 February 8 --- 12, 2005  Salt Lake City, Kolkata, India for their kind hospitality.
Dedicated to Professor  Bikash Sinha on occasion of his 60th anniversary. 
\vskip 0.3cm


\end{document}